\documentclass[lettersize,journal]{IEEEtran}
\usepackage{subfigure}
\usepackage{amsmath,amsfonts}
\usepackage{algorithmic}
\usepackage{algorithm}
\usepackage{array}
\usepackage[caption=false,font=normalsize,labelfont=sf,textfont=sf]{subfig}
\usepackage{textcomp}
\usepackage{stfloats}
\usepackage{url}
\usepackage{verbatim}
\usepackage{graphicx}
\usepackage{cite}
\begin{document}

\title{MobiWorld: World Models for Mobile Wireless Network}

\author{Haoye Chai,~\IEEEmembership{Member,~IEEE,} Yuan Yuan, Yong Li,~\IEEEmembership{Member,~IEEE}
\thanks{Haoye Chai is with the State Key Laboratory
of Networking and Switching Technology, Beijing University of Posts and Telecommunications, Beijing, China.

Yuan Yuan and Yong Li are with the Department of Electronic Engineering, Beijing National Research Center for Information Science and Technology (BNRist), Tsinghua University, Beijing 100084, China.}
}

\markboth{Journal of \LaTeX\ Class Files,~Vol.~14, No.~8, August~2021}%
{Shell \MakeLowercase{\textit{et al.}}: A Sample Article Using IEEEtran.cls for IEEE Journals}


\maketitle

\begin{abstract}
Accurate modeling and simulation of mobile networks are essential for enabling intelligent and cost-effective network optimization. In this paper, we propose MobiWorld, a generative world model designed to support high-fidelity and flexible environment simulation for mobile network planning and optimization.
Unlike traditional predictive models constrained by limited generalization capabilities, MobiWorld exhibits strong universality by integrating heterogeneous data sources, including sensors, mobile devices, and base stations, as well as multimodal data types such as sequences and images. It is capable of generating both network element-level observations (\emph{e.g.}, traffic load, user distribution) and system-level performance indicators (\emph{e.g.}, throughput, energy consumption) to support a wide range of planning and optimization tasks.
Built upon advanced diffusion models, MobiWorld offers powerful controllable generation capabilities by modeling the joint distribution between mobile network data and diverse conditional factors including spatio-temporal contexts, user behaviors, and optimization policies. This enables accurate simulation of dynamic network states under varying policy configurations, providing optimization agents with precise environmental feedback and facilitating effective decision-making without relying on costly real-network interactions.
We demonstrate the effectiveness of MobiWorld in a collaborative energy-saving scenario, where an agent uses observations and rewards generated by MobiWorld to optimize base station sleep and user offloading policies. Experimental results show that MobiWorld exhibits strong controllable generation performance and outperforms traditional methods in energy optimization.
\end{abstract}


\section{Introduction}
As mobile networks evolve towards 6G era, the complexity of network planning and optimization is rapidly increasing. Future networks will face significant challenges, driven by massive connectivity, diverse application scenarios, dynamic user behaviors, and extreme performance requirements. Network operators must navigate these complexities with intelligent decision-making processes, ensuring resource efficiency, user satisfaction, and seamless service delivery~\cite{10929033}. 

Central to effective network optimization is gathering accurate feedback from real-world comprehensive network performance under different conditions, like infrastructure configurations, user behaviors, and dynamic wireless channels~\cite{11018287}.
Such feedback forms the cornerstone for refining optimization policies iteratively, ultimately guiding networks toward optimal solutions. However, trial-and-error testing in live networks is impractical due to high costs, service disruption risks, and operational complexity. Consequently, network operators are increasingly inclined to leverage advanced digital twins and simulation technologies to enable safe and extensive experimentation within virtual environments. Yet, traditional rule-based or model-based simulations typically rely on overly simplified assumptions, which leads to a mismatch between the optimization solutions and real network behavior,  limiting their accuracy and reliability for precise and real-time optimization scenarios. 

Advances in data-driven deep learning models have shown strong potential in predicting mobile network metrics such as channel conditions and traffic load, and these models are extensively used to explore optimization policies in specific scenarios, such as base station placement, load balancing, and mobility management~\cite{10422716}.
Despite their successes, such specialized deep learning models possess inherent limitations. Specifically, these models focus narrowly on individual data types or single optimization scenarios, lacking the ability to learn and integrate multidimensional relationships in real mobile networks. More importantly, the predictive capabilities of these dedicated models are essentially limited to reproducing patterns observed in historical data, rather than generating differentiated network responses or feedback under varying optimization policies. 
The constraint significantly reduces their generalization capabilities, restricting their applicability across diverse and complex scenarios.

Recently, the concept of world models as generative frameworks based on foundation models has gained significant attention. From early advances such as Sora for video generation, to 3D environment modeling, and more recently 4D representations that capture dynamic spatio-temporal processes, world models have demonstrated a growing capacity to both understand complex data relationships and produce controllable outputs~\cite{nvidia2025cosmosworldfoundationmodel}. These capabilities have enabled their application across various fields, including autonomous driving, interactive gaming environments, and robotics, where they generate high-fidelity synthetic data to support simulation, planning, and real-time decision-making~\cite{yang2024learninginteractiverealworldsimulators, wang2023drivedreamerrealworlddrivenworldmodels}.
In the context of mobile networks, developing such world models holds substantial promise. 
On the one hand, world models can be primarily used to provide reliable and realistic environment feedback to allow agents to test different optimization policies and obtain accurate rewards, which is essential for effective optimization and decision-making, as seen in areas such as robotics and autonomous driving. This capability is exactly what mobile network planning and optimization require. It facilitates the construction of a safe and controllable virtual environment where control policy can be iteratively tested and refined without frequent interaction with the live network, thereby enhancing the reliability and robustness of final solutions.
On the other hand, mobile networks involve tens of thousands of interconnected devices that continuously generate massive data in real time, following standardized formats such as predefined frame structures and unified schemes, which provide a rich and reliable data foundation for training such world models. 
Despite its potential, there is still no consensus in the mobile networking community on what a world model entails or how to systematically design and implement it.

To bridge this research gap, we propose a world model paradigm for mobile networks, named MobiWorld, which is defined as a generative foundation model with powerful generalization capability that produces multiple high-fidelity network element-level data (\emph{e.g.,} cellular traffic, user mobility, application usage) and system-level performance data (throughput, coverage, energy consumption, \emph{etc.}), offering realistic feedback to guide planning and optimization.
MobiWorld provides a unified framework that models complex mobile networks through the integration of heterogeneous data sources such as sensors, user equipment, and base stations, and supports various data modalities, including sequential, tabular, and visual data.
Building with the advanced diffusion model, MobiWorld captures the joint distribution between mobile data and key conditioning factors, such as spatio-temporal contexts, user behaviors, and mobile infrastructure configurations. This allows for accurate simulation of network dynamics under different optimization policies, enabling agents to make informed decisions without direct reliance on costly real-world experimentation.
To the best of our knowledge, this is the first attempt to define a world model for mobile networks.
Through this work, we aim to identify three fundamental questions: (i) What role should a world model play in mobile networks? (ii) What are the core capabilities of such a model? (iii) How to train the world model for this domain?

The rest of this paper is structured as follows. Section I introduces the research background and fundamental definition of the MobiWorld. Section II presents its role and interaction mechanisms during the network planning and optimization process. Section III identifies the model’s core generative capabilities and the planning and optimization scenarios it supports. Section IV describes the backbone architecture and training methodology of the MobiWorld. In Section V, we provide a MobiWorld-enabled energy-saving optimization case, along with corresponding evaluation results. Finally, we conclude the paper and discuss future research directions in Section VI.

\section{The Paradigm of MobiWorld}
\label{sec:MobiWorld}

\begin{figure}[t]
	\centering
	\includegraphics[width=.8\linewidth]{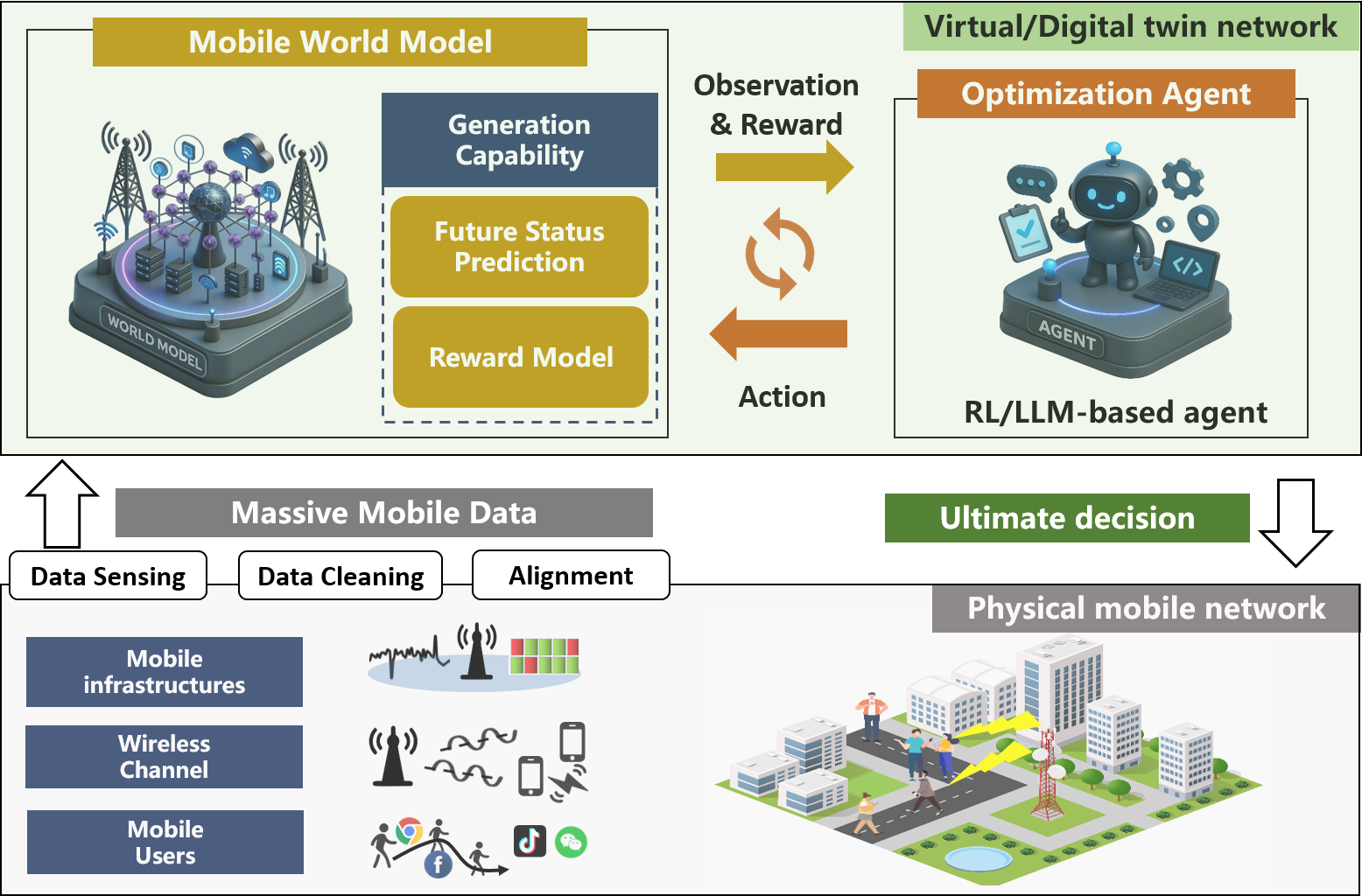}
\caption{The overview of MobiWorld and its role in mobile optimizations.}
\label{fig:role}
\end{figure}

The concept of world models can be traced back to the psychological notion of mental models introduced in 1971, where humans abstract the external world into simplified elements and achieve perception and cognition by simulating potential interaction outcomes~\cite{mentalworld}. Essentially, humans construct a mental/virtual replica of the external world, enabling accurate prediction of world-state changes under different conditions and actions. 
Current research into world models primarily follows two pathways: autoregressive methods and hierarchical planning-based Joint Embedding Predictive Architecture (JEPA) methods.
Both approaches emphasize the importance of enabling deep learning models to predict future data and provide accurate rewards for decision-making. From this perspective, the primary role of a world model in mobile networks is to \emph{offer high-fidelity, flexible, and generalizable environment feedback in virtual/digital twin to support agents in planning and optimization tasks}, as shown in Fig~\ref{fig:role}.

The mobile network world model can be conceptually understood as an AI-driven abstraction of the physical mobile network. It exists entirely in a virtual space and is designed to generate diverse types of network-related data that accurately reflect the dynamics of real-world mobile systems. By learning from large-scale, multi-modal datasets, the model captures the underlying correlations between infrastructure, user behavior, wireless conditions, and system-level performance, forming a high-fidelity simulated digital twin.

The primary function of the MobiWorld is to provide environment feedback, which manifests through its controllable generation capabilities. This feedback takes two main forms. The first concerns the generation of network element status, such as cellular traffic at base stations/cells, PRB (Physical Resource Block) utilization, application usage patterns, and mobility trajectories. These types of data can be used as observations during the network planning and optimization process. The second form involves generating system-level performance metrics, including Quality of Service (QoS) indicators such as network throughput, coverage, and overall energy consumption, and Quality of Experience (QoE) indicators, including user-perceived data rate, channel quality, and end-to-end latency, \emph{etc}. These performance-related outputs act as rewards to guide the optimization agent to explore policy.

\emph{MobiWorld-enabled optimization}. In the physical world, raw data collected from infrastructure components, user activity, and radio channel measurements are first processed through sensing, data cleaning, and data alignment pipelines. These data, encompassing both network element states and system performance metrics, are used to pretrain the MobiWorld. 
The pre-training process enables MobiWorld to effectively learn complex dependencies within large-scale mobile network data, such as the statistical relationships between group-level and individual-level attributes, as well as the underlying correlations between system performance metrics and network element states.
Once trained, MobiWorld is capable of generating network element data and system-level performance metrics that closely align with real-world mobile network conditions. These two types of data serve as environmental observations and reward signals for optimization agents, which can be grounded in reinforcement learning or large language models (LLMs) paradigms.
The optimization agents iteratively refine their policies based on feedback from the MobiWorld and feed the updated policies back for further evaluation. This closed-loop process enables efficient trial-and-error optimization within the virtual digital twins, reducing reliance on costly real-world experiments. Once a policy converges through repeated iterations, the ultimate solutions can be reliably deployed in the physical mobile network.

\section{Core Capabilities of MobiWorld: Controllable Generation}
\label{sec:core}

\begin{figure}[t]
	\centering
	\includegraphics[width=.8\linewidth]{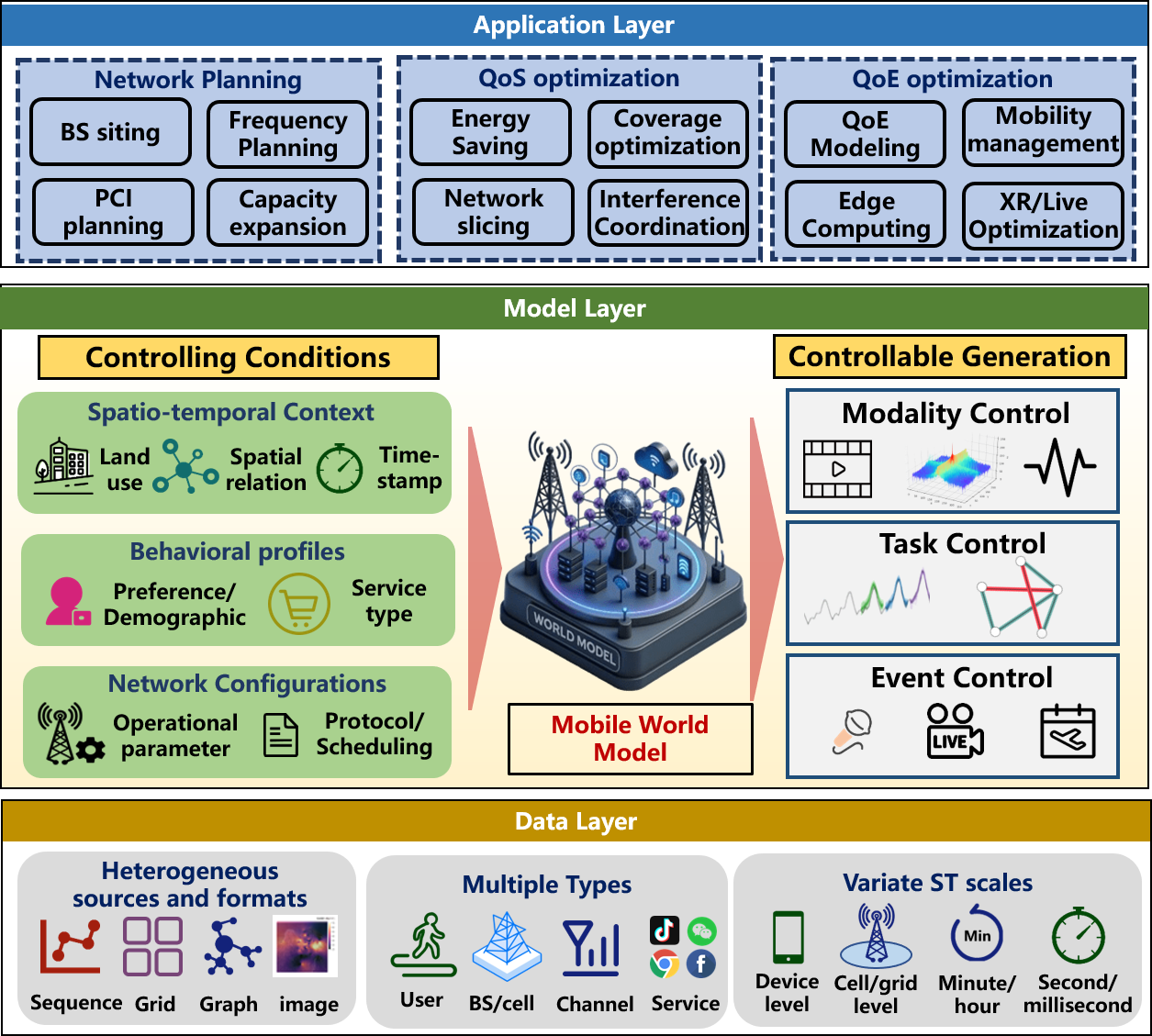}
\caption{The system architecture of the MobiWorld.}
\label{fig:enabler}
\end{figure}

In the MobiWorld-enabled optimization process, the optimization agents continuously propose new policies and interact with the MobiWorld to receive environment feedback for refinement.
At the heart of this iterative trial-and-error loop lies the \emph{controllable generation capability}.
Unlike small conventional deep learning models, which typically focus on predicting future values by learning deterministic patterns from historical data (such as consistent traffic patterns observed in fixed spatio-temporal contexts), the MobiWorld must operate under more dynamic and counterfactual conditions. Since policy exploration often leads to previously unseen configurations, including rare environment states and unobserved resource deployments. Under such circumstances, historical prediction alone becomes insufficient. Instead, the model must be capable of generating mobile network data that reflects these newly controlling policies. 
To clarify the controllable generation capability, we propose a framework for the MobiWorld, which consists of three layers, as shown in Fig~\ref{fig:enabler}.

\emph{Data layer.}
This layer serves as the foundational input for MobiWorld.
To enable robust and controllable generation, it is essential to collect data with sufficient diversity, allowing the model to explore a wide range of scenario-specific characteristics. This diversity is reflected in three key aspects: the heterogeneity of data sources, such as network infrastructure, mobile devices, and sensors; the variety of data types, including user trajectories, wireless channel measurements, and service usage logs; and the inclusion of multiple spatio-temporal scales, ranging from cell-level, grid-level, and individual user-level spatial granularity to temporal resolutions spanning from hours to milliseconds.


\emph{Model layer.} The layer is responsible for integrating and learning from large-scale, heterogeneous datasets to enable controllable data generation. The core idea behind controllable generation lies in modeling the joint distribution between mobile network data and a set of controlling conditions. Given a set of conditions, the model is able to sample mobile network data from the learned joint distribution that is consistent with those conditions, thereby simulating realistic network responses under diverse scenarios.
These controlling conditions can be broadly categorized into three domains:

$\bullet$ Spatio-temporal context, including features such as land use, POI (Point of Interest) distribution, and timestamps, enables the model to capture geographic and temporal variations in mobile network patterns. 
This category primarily helps the model learn the patterns and dynamics of mobile data variations across different regions and periods.

$\bullet$ Behavioral profiles, which encompass user preferences, application types, and service metadata, are used to characterize behavioral patterns under different usage contexts. This category helps uncover the latent dependencies between user behavior and mobile data dynamics.

$\bullet$ Network configurations, such as base station operational parameters and resource scheduling policies, play a crucial role in connecting optimization policies to network performance. 
Compared to spatio-temporal environmental factors or user behavior data, agents in network planning and optimization are more likely to adjust network configurations, such as base station transmission power and antenna tilt, to search for optimal solutions. Establishing a strong coupling between policies and network configurations allows MobiWorld to generate aligned network dynamics under different exploration policies, thereby providing agents with accurate and consistent rewards for effective learning.

From a structural perspective, MobiWorld is essentially a generative foundation model. This foundational model is designed to handle high-dimensional, multi-modal data with complex, non-linear distributions that characterize the behavior of mobile networks. By leveraging large-scale training across diverse scenarios, it can accurately learn the joint distributions between mobile data and controlling conditions.
On the output side, MobiWorld is expected to support three dimensions of controllable generation:

$\bullet$ Modality controllability: The MobiWorld can generate a variety of data modalities depending on the optimization tasks. This includes time-series data that reflect temporal variations in mobile networks, frequency-domain data like spectrograms to characterize wireless signal properties, and vision data such as signal strength heatmaps or user distribution maps that capture spatial patterns.
This modality flexibility is essential for supporting different downstream applications, such as temporal forecasting and spectrum sensing.

$\bullet$ Task controllability: Beyond data modality, MobiWorld supports various core tasks commonly encountered in mobile network analysis and optimization. These include time-series forecasting, which is fundamental for predicting traffic loads and resource demand; graph completion, which is important for inferring missing relationships in spatial or topological structures such as cell adjacency graphs or user mobility networks; and image generation tasks, which can be used to simulate signal coverage properties. By conditioning on different task-specific inputs, the model can adapt its generative behavior to fulfill the objective of each optimization scenario, making it compatible with multi-task learning pipelines.

$\bullet$ Event controllability: The model is expected to simulate the impact of rare but high-impact events, such as concerts, public gatherings, on mobile network data. These events often lead to dramatic, short-term shifts in user density, traffic volume, and service demand, which traditional models struggle to capture due to their long-tail distribution in historical data. This requires the model to possess strong few-shot learning capabilities and a deep understanding of how external events influence user behavior and network dynamics.

\emph{Application layer.} The layer leverages the controllable generation capabilities of the model to enable a wide range of planning and optimization tasks in virtual space. These tasks can be broadly categorized into three types:

$\bullet$ Planning-oriented applications rely on the model’s capacity to generate mobile data under specific urban spatio-temporal conditions. By simulating large-scale network performance under specific land use or base station deployment conditions, the model supports scenario-driven planning decisions, such as estimating network demand and finding the optimal cell siting solutions across different urban layouts.

$\bullet$ QoS-oriented applications focus on system-level performance metrics under various network infrastructure configurations~\cite{10325531}. These tasks exploit the model’s capacity to generate mobile network data in response to changes in operational parameters or resource scheduling schemes. For example, the model can simulate the number of active users and the power consumption of cells under different carrier shutdown schemes, providing insights for energy-efficient optimization.

$\bullet$ QoE-oriented applications emphasize user-centric optimization by modeling subjective experience metrics~\cite{9877931}. These applications use conditions derived from user behavior preferences or service types to simulate end-to-end quality indicators such as latency, throughput, \emph{etc}. The ability to generate realistic user-level performance data enables fine-grained policies and improvement of perceived service quality.

\section{Backbone of MobiWorld: Foundation Model}
\label{sec:enabler}

\begin{figure}[t]
	\centering
	\includegraphics[width=\linewidth]{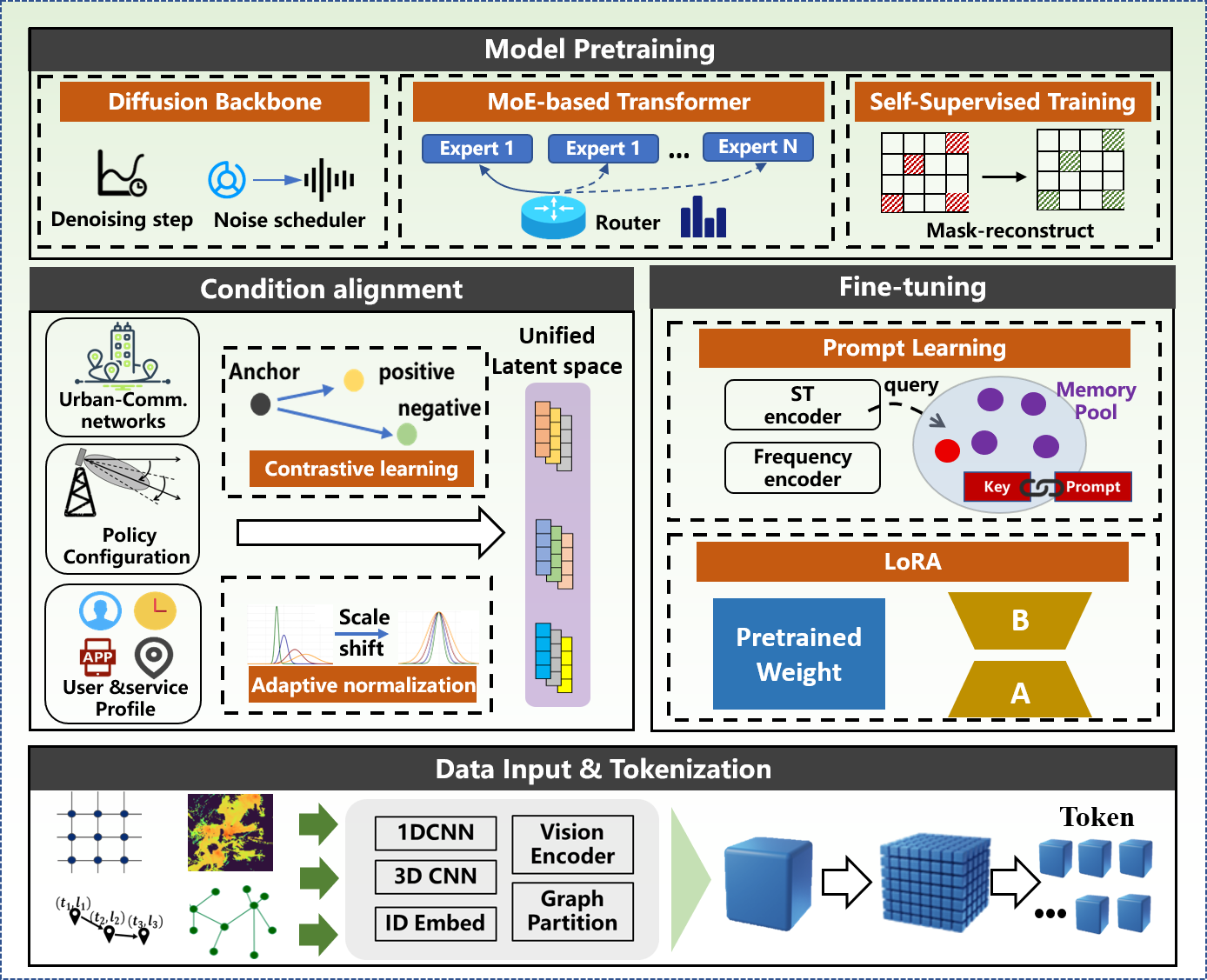}
\caption{Foundation model as the backbone of the MobiWorld. The framework comprises four key components: data input and tokenization, condition alignment, pretraining, and fine-tuning.}
\label{fig:backbone}
\end{figure}

To enable controllable generation in MobiWorld, the design of its backbone architecture plays a crucial role.
Motivated by the recent success of diffusion models in natural language and computer vision tasks, which have been proven highly effective in capturing complex data distributions and enabling conditional generation~\cite{10.5555/3540261.3540933}, the MobiWorld is constructed with a diffusion-based generative model. As shown in Fig~\ref{fig:backbone}, the framework comprises four key components: data tokenization, condition alignment, pretraining, and fine-tuning.

\emph{Data Input and tokenization.} It serves to unify the representation of heterogeneous mobile network data, such as sequences, images, and graphs, into a consistent, model-compatible format. This process involves transforming data with diverse structures and spatio-temporal scales into a shared low-dimensional token space. Specifically, we apply techniques such as convolutional encoders for time-series and spatial grids, graph partitioning methods for topological data, and vision encoders for image-based inputs.  
These representations are then segmented into a sequence of tokens, which serve as standardized inputs for model training.

\emph{Condition alignment}. It involves mapping diverse controlling conditions into a unified latent vector space, allowing the model to interpret and utilize them consistently during generation. However, these conditions often differ significantly in their semantic meaning; for example, a timestamp encodes temporal periodicity, while a configuration parameter like transmission power reflects system-level intent. Such semantic differences are typically accompanied by distinct distributional characteristics, making direct integration challenging. 
One effective approach is contrastive learning, which encourages semantically related conditions to have similar representations in the latent space, thereby reducing the semantic gap between heterogeneous inputs. In addition, normalization methods also help to mitigate differences in scale and distribution across condition types~\cite{10377858}, ensuring convergence during training and generation stability.

\emph{Model Pretraining}. During the pretraining phase, the self-supervised learning strategy can be leveraged with masking-reconstruction mechanisms, applied to the backbone architecture that combines diffusion models with Transformer structures. It enables the model to learn robust representations by reconstructing masked portions of the input data, thereby capturing both the underlying structure and implicit dependencies within mobile network data.
To capture multi-scale features in mobile data, Mix-of-Experts (MoE) Transformers~\cite{10.5555/3586589.3586709} can be utilized comprising a gating network $g_\rho(x_t)$ and multiple expert networks $ m_{\theta_i}(x_t)$. Each expert focuses on learning a specific type of high-dimensional feature, while the gating network assigns weights or probabilities to different experts based on the prompt and input features, which yields
\begin{equation}
    y(x_t, \theta_1, \theta_2, ... \theta_M; \rho)  = \sum_{i=1}^M g_\rho(x_t) m_{\theta_i}(x_t).
\end{equation}
The mechanism enables the MoE to disentangle complex high-dimensional representations and facilitates dynamic expert collaboration, enhancing the model’s generative capacity.

\emph{Fine-tuning}. This module is essential for adapting the pretrained MobiWorld to diverse downstream optimization tasks while retaining its generalizable knowledge. To this end, the Low-Rank Adaptation (LoRA) can be employed yo insert trainable low-rank matrices into specific layers, enabling efficient parameter updates with minimal overhead. This approach supports task-specific adaptation without compromising previously learned representations. In addition, we incorporate prompt learning to guide the model’s attention to salient data features, improving its capacity to model the joint distribution between control conditions and mobile network data~\cite{10.1145/3560815}.
For example, a memory network can be leveraged to store and retrieve multi-dimensional knowledge in the data (\emph{e.g.,} periodicity, frequency pattern), which can be denoted as a set of key-prompt pairs built on learnable parameters, $\{(k_1, P_1), (k_2, P_2), ..., (k_N, P_N)\}$.
Each encoded pattern $x$ serves as a query $q_x$ to index the memory pool, retrieving the most similar set of keys $\mathbf{K}_x = \{k_i\}_{i=1:n}$. The corresponding parameters $\mathbf{P}_x$ can be regarded as prompts and fed into the training process.
This approach enables the prompt to automatically explore patterns during model training and retrieve the most relevant information to navigate the forecasting process.

\section{Case study: MobiWorld-enabled Energy Saving}
\label{sec:usecase}

\begin{figure}[t]
	\centering
	\includegraphics[width=.7\linewidth]{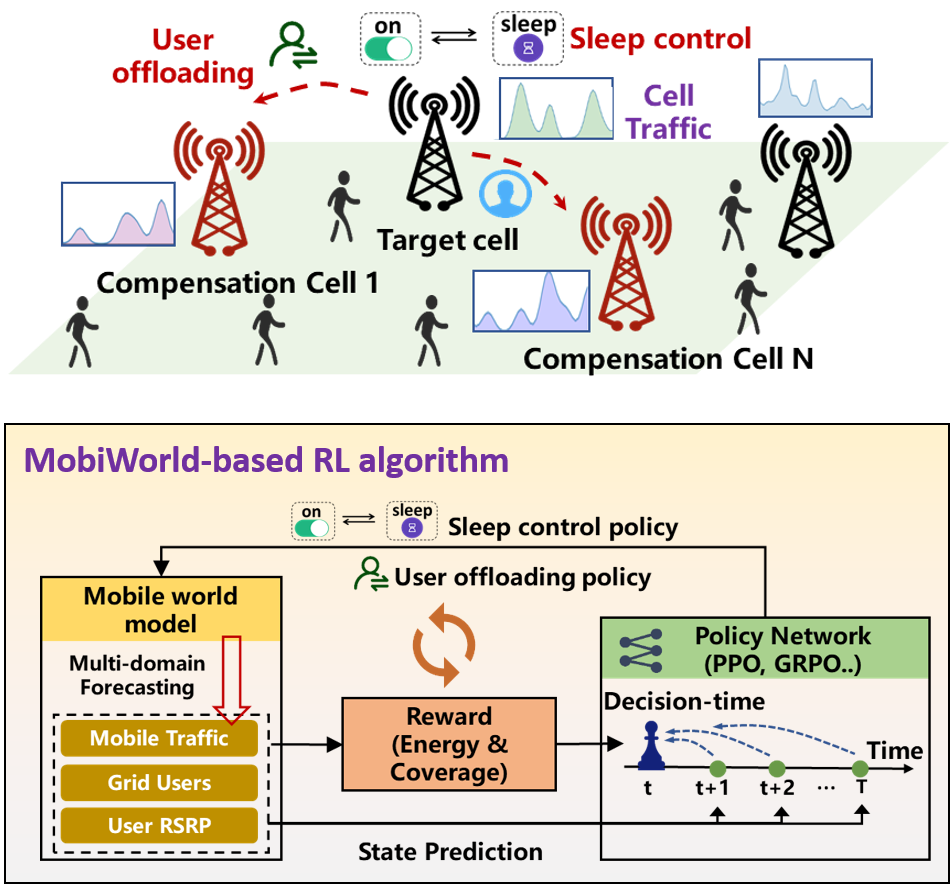}
\caption{MobiWorld-enabled energy saving optimization. (Top: Multi-cell cooperative energy-saving scenario, where cells control sleep states and user offloading to reduce energy consumption. Bottom: Interaction between MobiWorld and the optimization agent.)}
\label{fig:usecase}
\end{figure}

This section presents a MobiWorld-enabled energy optimization case to illustrate the model’s effectiveness.

\subsection{Problem description}

As shown in Figure~\ref{fig:usecase}, we consider a multi-cell collaborative energy-saving problem in mobile networks. Each cell can control sleep status by switching carriers on or off to improve energy efficiency during low-demand periods. Each cell is also linked to a set of compensation neighboring cells that can offload users through modified association mechanisms, including cell selection bias, handover threshold adjustment, and load-aware strategies.
The objective is to minimize total network energy consumption while maintaining user experience, quantified by the average Reference Signal Received Power (RSRP) across all users. Each cell must determine its sleep mode and the number of users to offload to neighboring cells.

Solving this problem requires two types of data. The first is network element states, including cellular traffic and the number of users within each cell’s coverage grid where The former reflects the load condition used to guide cell sleep control policy, and the latter determines the available user numbers to be offloaded to neighboring cells. The second is user experience data, specifically device-level RSRP measurements, which can be used to evaluate network-level coverage performance.



The MobiWorld leverages its capability for complex data modeling and controllable generation to produce mobile data that not only resembles real-world conditions but also adapts dynamically to different base station operational configurations.  
The controllable generation manifests in two key aspects. First, given the geographic context and temporal features of each cell and its surrounding grids, the model can generate spatio-temporally aligned cellular traffic data and user distribution patterns that reflect realistic environmental dynamics. Second, by incorporating base station frequency configurations and conditioning on specific sleep control policies during each optimization round, MobiWorld can generate corresponding RSRP data that reflect users' perceived signal strength under different energy-saving decisions. This dual capability enables MobiWorld to act as a flexible environment simulator, supporting policy-aware and fine-grained network optimization.

The MobiWorld-based optimization process can be described as follows. MobiWorld serves as a controllable environment simulator, producing traffic load and user distribution data as environmental observations, and generating user-level RSRP values as part of the optimization rewards. These generated data are passed to an optimization agent, which can be PPO, MAPPO, or other RL schemes.
The agent incorporates not only the current network state but also the predicted future traffic load and user distribution for both the target cell and its compensation neighbors. Based on this information, it explores energy-saving policies by maximizing a reward function that jointly accounts for total network energy consumption and average user RSRP.
At each iteration, the agent produces a joint action that includes sleep policy and user offloading decisions. These candidate policies are then fed back into MobiWorld, which updates the simulated environment accordingly. Through repeated closed-loop interaction, the optimization module progressively converges to an energy-efficient policy that effectively balances energy savings with user experience.

\subsection{Performance Evaluation}


\begin{figure}[t]
    \centering
    \subfigure[Cell traffic and Grid Users.]
    {\includegraphics[width=.47\linewidth]{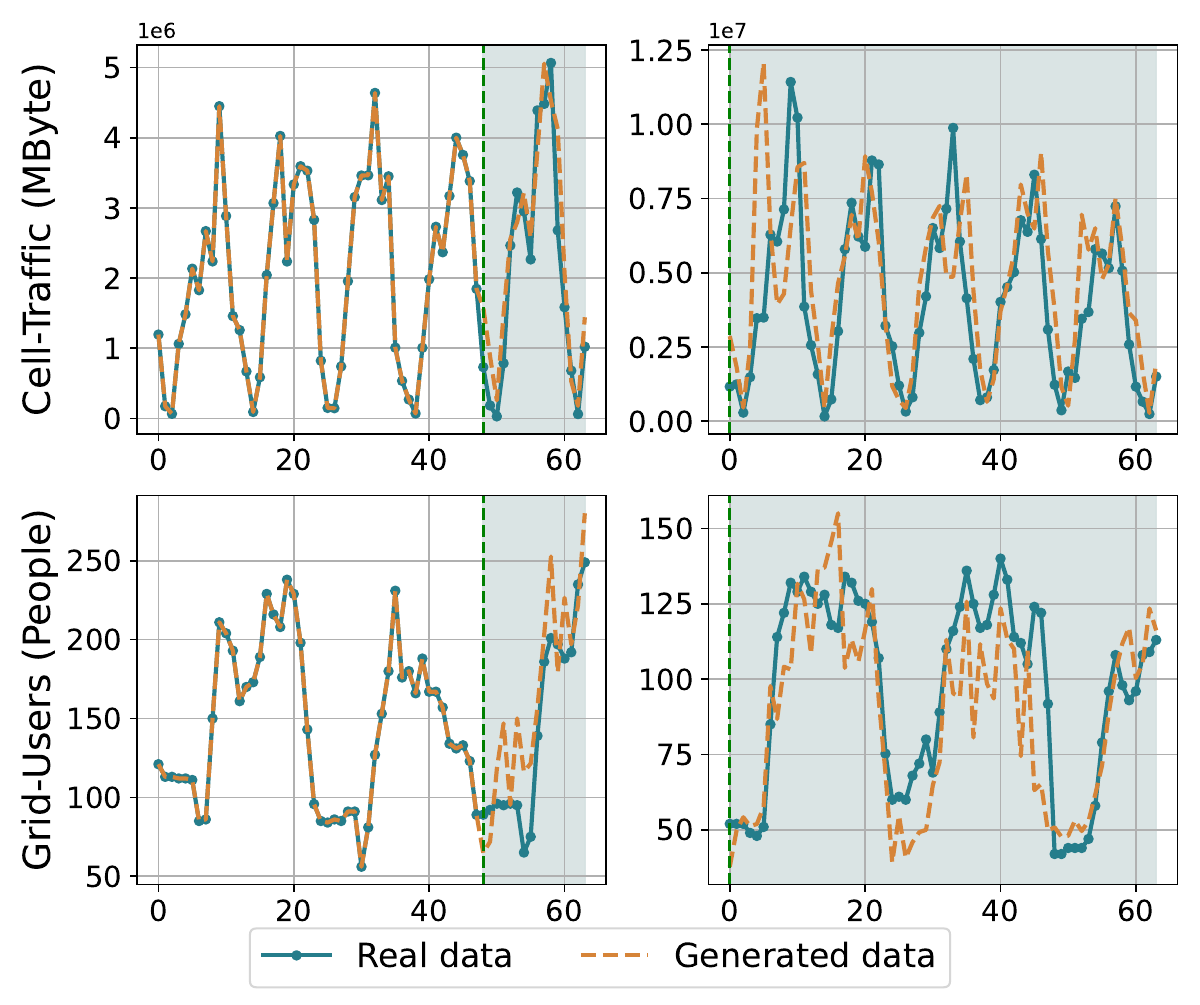}}
    \subfigure[User RSRP.]{\includegraphics[width=.47\linewidth]{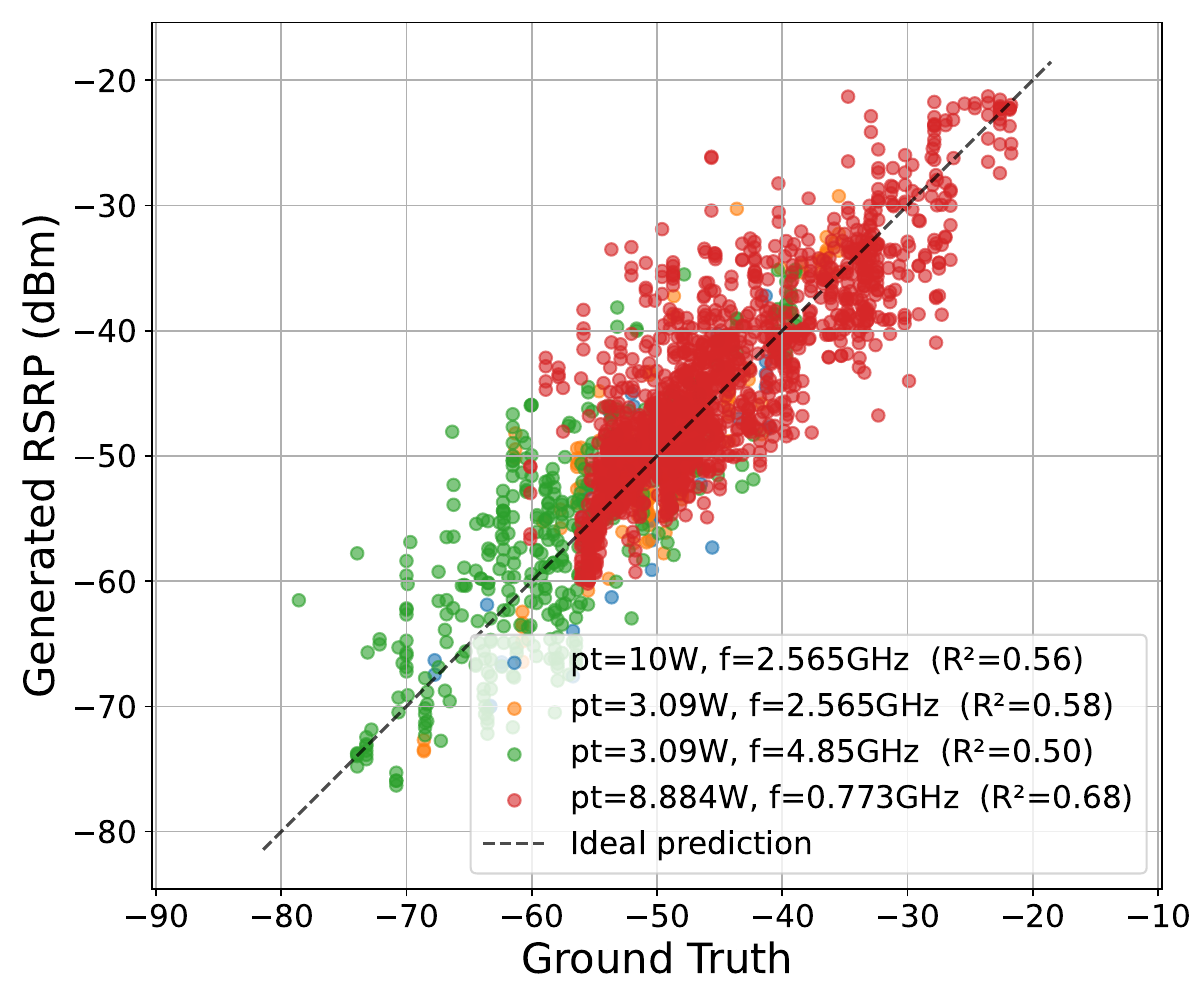}}
    \caption{Controllable generation results. ( Left: Traffic/User generation based on the spatio-temporal context of cell/grid. Right: Controllable user-side RSRP generation  conditioned on base station operating parameters.)}
    \label{visual_res}
\end{figure}



We evaluate the performance of the proposed MobiWorld in two key aspects: controllable data generation and effectiveness of optimization.
Figure~\ref{visual_res}(a) presents the generation results for cell traffic (at a 2-hour granularity) and grid-level user count (at a 1-hour granularity). To demonstrate MobiWorld’s generation capability, we consider two distinct tasks. The left column shows the short-term prediction task, where the model generates traffic and user data for the near future (highlighted in green) conditioned on both historical data and the surrounding spatio-temporal context of the cell. In contrast, the right column shows the long-term generation task, where the model produces mobile data solely based on environmental features, without access to historical inputs. As shown, MobiWorld accurately generates both traffic and user distributions under different conditions, demonstrating high-fidelity generation capabilities.

Figure 1~\ref{visual_res}(b) illustrates the generated user RSRP values under different cell energy-saving policies. The x-axis represents the ground-truth values, while the y-axis denotes the corresponding model-generated outputs. In this scenario, MobiWorld generates RSRP values under conditions including cell transmit power, carrier frequency, and user-to-cell distance, aiming to evaluate its controllability under varying energy-saving policies, \emph{i.e.}, after a cell enters sleep mode, the user-side RSRP is generated based on the operating parameters of the new cell they connect to. The results show that MobiWorld can flexibly generate RSRP under different operating parameters, providing a solid foundation for optimization agents.

\begin{figure}[t]
	\centering
	\includegraphics[width=.8\linewidth]{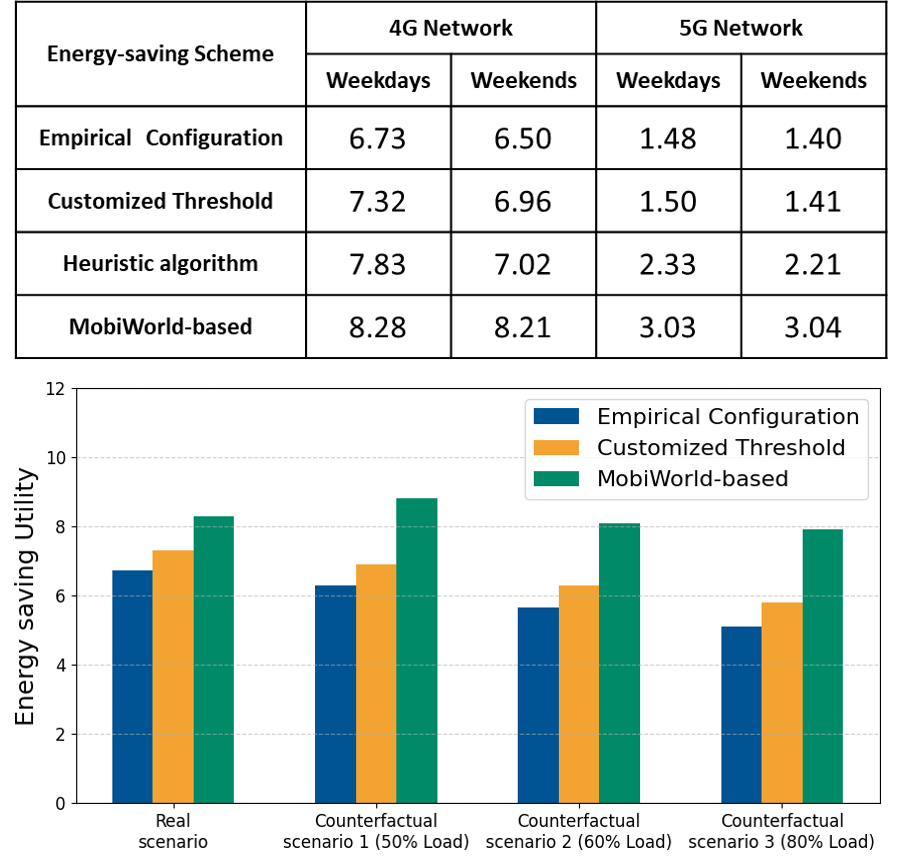}
\caption{Energy optimization results. The top figure shows the energy saving utility achieved by different schemes across various cell types and time periods. The bottom figure presents the energy saving utility under counterfactual scenarios with varying levels of cellular traffic load. The energy saving utility is computed as a weighted sum of total cell energy consumption and the average RSRP across all users.}
\label{fig:res2}
\end{figure}

For energy-saving optimization, we evaluate three baseline algorithms: empirical configuration, customized thresholding, and a heuristic method. The first two rely on predefined traffic thresholds to trigger energy-saving actions~\cite{5978418}, while the heuristic scheme relies on a greedy rule-based algorithm that iteratively shuts down cells within a region based on
a fixed priority order~\cite{7818423}. As shown in the upper part of Figure~\ref{fig:res2}, our method consistently outperforms the baselines across different scenarios. This improvement can be attributed to the fact that baseline methods primarily consider cell load status and lack the ability as the MobiWorld to comprehensively simulate multi-dimensional network environments. including traffic volume, user distribution, and RSRP. As a result, their policy decisions often fall into local optima, limiting overall performance.

Moreover, in real-world network operations, most cells typically operate under idle or low-load conditions. As a result, conventional optimization methods that rely solely on live network data are inherently biased; in other words, the policy often lacks exposure to high-load or bursty scenarios, which limits their effectiveness in handling rare but critical scenarios.
MobiWorld, with its controllable generation capability, addresses this limitation by generating large-scale, high-fidelity counterfactual data under diverse load and configuration settings. These synthetic scenarios enable robust counterfactual reasoning and strategy evaluation.
To validate this, we conduct additional experiments under three counterfactual high-load scenarios, where the traffic load of a given cell reaches 50\%, 60\%, and 80\% of its capacity, respectively. As shown in the lower part of Fig~\ref{fig:res2}, our method consistently achieves strong optimization performance across all counterfactual scenarios, demonstrating superior robustness and generalization of MobiWorld-enabled optimization.

\emph{Summary}. MobiWorld offers two key advantages over traditional optimization approaches. First, its pretraining process jointly models multi-dimensional data, enabling it to capture complex dependencies and simultaneously generate rich mobile network data. Second, its controllable generation capability allows the model to simulate system-level variations under different configurations and to synthesize large volumes of high-fidelity data for counterfactual scenarios, thereby improving the policy exploration and optimization effectiveness.

\section{Conclusion and future directions}
\label{sec:con}

In this paper, we introduced MobiWorld,  a controllable generative world model for mobile network planning and optimization. By learning from massive and complex real-world data, MobiWorld enables high-fidelity and flexible generation across different data modalities, formats, and spatio-temporal scales.
The model also possesses strong controllable generation capabilities that enable it to simulate both network element-level status and system-level performance metrics under varying configurations and policy conditions.
Evaluations on an energy-saving case demonstrate that MobiWorld offers an effective and low-cost solution that enables decision-making in virtual or digital twin environments and supports high-fidelity data generation under counterfactual scenarios.
In addition to the advantages discussed above, many challenges remain across data, model design, and practical applications. We outline several potential future directions below.

$\bullet$  Enhancing comprehensive controllability: Future work may explore more flexible and fine-grained conditioning mechanisms to enable the world model to adapt to a broader range of scenarios. This includes the integration of richer semantic inputs, such as policy intents, network topology variations, or user mobility patterns, to further improve the model’s interpretability and control precision.

$\bullet$  Fusing multi-modal mobile network data: A key direction is to unify data representations across time, frequency, and pixel domains. Building cross-domain modeling capabilities can help the world model bridge time-series dynamics, spectral characteristics, and spatial structures, enabling more holistic and high-resolution simulation of real-world mobile network behavior.


$\bullet$ Large-Scale Deployment:
Future research should focus on applying the MobiWorld to large-scale networks with diverse service demands and complex data distributions. A key challenge is how to support multi-dimensional task objectives, such as joint optimization of load balancing, energy efficiency, and latency reduction. This requires tightly integrating world models with intelligent agents for closed-loop optimization. The model must generate controllable, task-aware data in response to varying network conditions. Additionally, improving scalability and inference efficiency is essential for real-time deployment. Techniques like model compression, modular design, and efficient sampling will be critical for enabling practical use in large-scale mobile network environments.

\bibliographystyle{IEEEtran}
\bibliography{Reference}

\vfill

\end{document}